\documentclass{article}
\usepackage{spconf,amsmath,graphicx}

\usepackage{graphicx}
\usepackage{subfigure}
\usepackage{booktabs}
\usepackage{cite}
\usepackage{microtype}
\usepackage{xurl}

\usepackage{amsmath}
\usepackage{amssymb}
\usepackage{mathtools}
\usepackage{amsthm}
\usepackage{textcmds}
\usepackage{color}
\usepackage[flushmargin]{footmisc}
\usepackage{arydshln}
\usepackage{comment}

\definecolor{azure(colorwheel)}{rgb}{0.0, 0.5, 1.0}
\definecolor{nicegreen}{rgb}{0.0, 0.7, 0.1}
\definecolor{CuGray}{gray}{0.9}
\definecolor{pink}{cmyk}{0, 0.7808, 0.4429, 0.1412}
\definecolor{amethyst}{rgb}{0.6, 0.4, 0.8}
\definecolor{black}{rgb}{0.0, 0.0, 0.0}
\definecolor{purple}{rgb}{0.6275, 0.0275, 0.6706}

\definecolor{steelblue}{rgb}{0.27, 0.51, 0.7}
\definecolor{brickred}{rgb}{0.8, 0.25, 0.33}

\definecolor{customgray}{rgb}{0.9, 0.9, 0.9}

\renewcommand{\paragraph}[1]{\vspace{1mm}\noindent\textbf{#1.}\,\,\,}

\usepackage[breaklinks,colorlinks,citecolor=nicegreen,linkcolor=nicegreen,bookmarks=false]{hyperref}
\usepackage[capitalize,noabbrev]{cleveref}
\usepackage{enumitem}
\usepackage{multirow}

\title{Unsupervised Pre-training for Data-Efficient Text-to-Speech \\ on Low Resource Languages}
\name{Seongyeon Park$^{1\ast}$, Myungseo Song$^{1\ast}$, Bohyung Kim$^{1}$ and Tae-Hyun Oh$^{2,3}$
\thanks{$^\ast$Equal contribution}}
\address{$^{1}$CNAI, Seoul, Korea \\ 
$^{2}$Dept. of EE and GSAI, POSTECH, Pohang, Korea\\
$^{3}$Institute for Convergence Research and Education in Advanced Technology,\\
Yonsei University, Seoul, Korea}

\begin{document}
\ninept
\maketitle

\def\eg{\emph{e.g.}}
\def\Eg{\emph{E.g.}}
\def\ie{\emph{i.e.}}
\def\Ie{\emph{I.e.}}
\def\etal{et al.}

\newcommand{\m}{\mathbf{m}}

% !TEX root = ../main.tex

\begin{abstract}
Neural text-to-speech (TTS) models can synthesize natural human speech when trained on large amounts of transcribed speech.
However, collecting such large-scale transcribed data is expensive.
This paper proposes an unsupervised pre-training method for a sequence-to-sequence TTS model by leveraging large untranscribed speech data.
With our pre-training, we can remarkably reduce the amount of paired transcribed data required to train the model for the target downstream TTS task. 
The main idea is to pre-train the model to reconstruct de-warped mel-spectrograms from warped ones, which may allow the model to learn proper temporal assignment relation between input and output sequences.
In addition, we propose a data augmentation method that further improves the data efficiency in fine-tuning.
We empirically demonstrate the effectiveness of our proposed method in low-resource language scenarios, achieving outstanding performance compared to competing methods.
The code and audio samples are available at: \url{https://github.com/cnaigithub/SpeechDewarping}
\end{abstract}

% !TEX root = ../main.tex

\section{Introduction}
\label{sec:introduction}
Recent advance in deep neural networks enables us to build end-to-end text-to-speech (TTS) models~\cite{wang2017tacotron, shen2018natural} to synthesize plausible speech.
Recent research~\cite{chung2019semi, zhang2020unsupervised} attributes natural and plausible speech generation of TTS models to the following capabilities to be learned: 1) \emph{attention alignment} between the input and output sequences, and 2) \emph{autoregressive prediction} of acoustic features.
The supervised learning or pre-training methods~\cite{chen2018sample,moss2020boffin} directly inject the necessary capabilities for TTS through supervision using large-scale transcribed speech. 
However, such models require a large amount of transcribed speech data for training, which is not annotation efficient.
Constructing such large-scale text-annotated speech is time-consuming, costly, and even infeasible for low-resource languages.

To mitigate the labeled data deficiency, pre-training methods for TTS systems have been investigated~\cite{chen2018sample,moss2020boffin,chung2019semi,zhang2020unsupervised}.
Among them, Chung~\etal and Zhang~\etal\cite{chung2019semi, zhang2020unsupervised} specifically designed to induce either of such capabilities in unsupervised ways by leveraging large-scale untranscribed speech data.
In \cite{chung2019semi}, the decoder of Tacotron~\cite{wang2017tacotron} is pre-trained as an autoregressive speech generator.
In \cite{zhang2020unsupervised}, the whole model of Tacotron~2~\cite{shen2018natural} is pre-trained to predict speech from unsupervised linguistic units extracted by an external Vector-quantization Variational-Autoencoder (VQ-VAE)~\cite{chorowski2019unsupervised}.
It would be desirable to pre-train the full TTS model without any external model.

The goal of this paper is to further reduce the amount of transcribed speech required for TTS training.
To this end, we propose an unsupervised pre-training method for Tacotron~2, \emph{Speech De-warping}.
By utilizing large-scale \emph{untranscribed} speech, our key idea is to make the TTS model learn to reconstruct original spectrograms from warped ones, \ie, learn to \emph{de-warp}.
This method does not require annotation, as we synthesize the warped spectrograms by a simple random temporal warping technique.
We sample random segment boundaries and resize
each segment along the temporal axis to be a fixed size.
Learning to de-warp as a pre-training step encourages the model to acquire both preliminary knowledge of attention alignment and autoregressive prediction.
After the pre-training, we fine-tune the model using small-scale transcribed speech data of a target speaker, possibly in a low-resource language.
In addition, we extend our simple random warping technique to a data augmentation method for the fine-tuning step, which further improves performance.

Compared to the previous studies, our pre-training method does not suffer from the model mismatch problem between pre-training and fine-tuning~\cite{chung2019semi} and does not require training an external model for data preparation~\cite{zhang2020unsupervised}.
It is also worth noting that our data augmentation does not require any external data or pre-trained models unlike other data augmentation approaches for TTS~\cite{hwang2021tts,song2022tts,oh2022effective,comini2022low,terashima2022cross,huybrechts2021low};
they typically leverage a large amount of transcribed speech to generate synthetic data with pre-trained TTS models or voice conversion models.

Our main contributions are summarized as follows:
1) proposing an unsupervised pre-training method for TTS models, \emph{Speech De-warping},
2) proposing a simple yet effective data augmentation method, \emph{SegAug},
3) demonstrating improved data efficiency,
and 4) showing the cross-language effectiveness of our methods.
% !TEX root = ../main.tex

\section{Proposed Method}
\label{sec:method}

\subsection{Segment-based Speech Warping}
Our pre-training and data augmentation methods include the procedure of warping speech.
To warp the speech, we segment the mel-spectrograms of the speech along the time axis and apply a transformation per segment.
To clarify the procedure, we describe the general form of the segment-based speech warping $f$.

Given a mel-spectrogram $\m$ of timesteps $N$, we warp $\m$ to generate a warped mel-spectrogram $\hat{\m}$, which is given by
\begin{equation}
    \label{eqn:warping}
    \hat{\m} = f(\m; S, T),
\end{equation}
where $S$ is a segmentation method, and $T$ is a transformation.
The segmentation method $S$ segments $\m$ into $k$ different spectrogram segments $\m_{1}, \m_{2}, ..., \m_{k}$ such that $N = \sum_{i=1}^{k}N_{i}$, where $N_{i}$ is the number of timesteps of $\m_{i}$.
Then, for each segment $\m_{i}$, the transformation $T$ transforms $\m_{i}$ to a warped segment $\hat{\m_{i}} = T(\m_{i})$.
We concatenate the warped segments along the time axis to generate the warped spectrogram $\hat{\m} = concat(\hat{\m_{1}}, \hat{\m_{2}}, ..., \hat{\m_{k}})$.

Theoretically, any segmentation method and transformation can be used as $S$ and $T$, \eg, phoneme segmentation for $S$.
We present our specific configuration for $S$ and $T$ in the following subsections.

\subsection{Pre-training: Unsupervised Speech De-warping}
\begin{figure}[t]
% \vskip 0.2in
\centering
% \begin{center}
% \centerline{
\includegraphics[width=1\columnwidth]{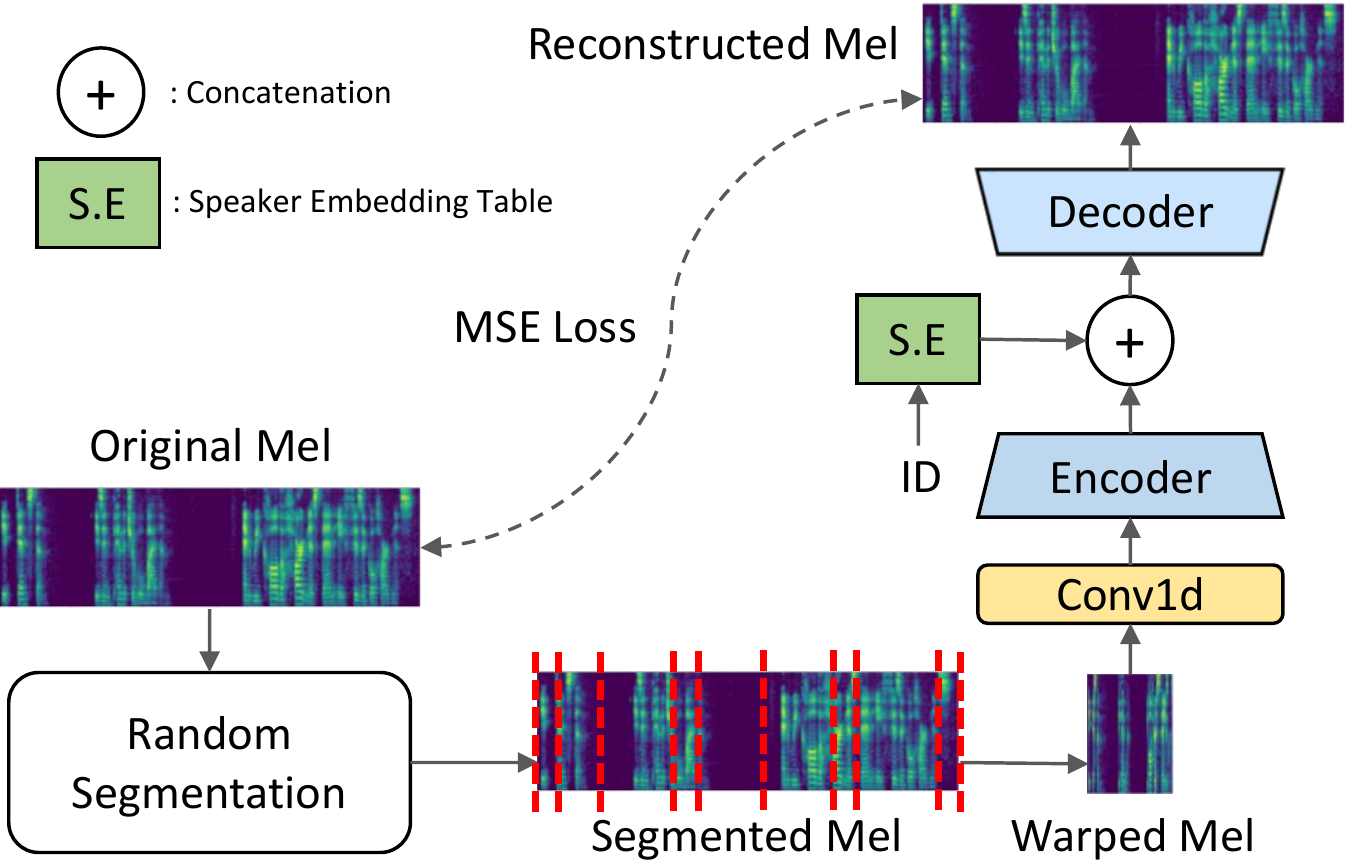}
% }
% \vspace{-3mm}
\caption{Overview of our
% Our unsupervised pre-training method, 
\emph{Speech De-warping}.
We randomly segment the spectrogram and warp it by resizing each segment to have equal unit timesteps.
% ,  \ie, 1.
The model is learned to reconstruct the original spectrogram from the warped one.
}
\label{fig:our_framework}
% \end{center}
% \vspace{-4mm}
\end{figure}

We aim to reduce the amount of transcribed speech required for TTS training.
To this end, we propose an unsupervised pre-training method, \emph{Speech De-warping} which leverages large-scale \emph{untranscribed} speech, which is much cheaper to obtain than transcribed one.
The main idea is to pre-train a TTS model to recover original spectrograms from warped ones, which is illustrated in Figure~\ref{fig:our_framework}.

To generate pairs of input and expected output for unsupervised learning, we first generate warped spectrograms from the original spectrograms converted from the untranscribed speech by the segment-based speech warping (see Equation~\ref{eqn:warping}).
Specifically, we use random segmentation as $S$, which randomly selects $k-1$ number of boundary timesteps; thus, 
$k$ segments are obtained.
The segment boundaries are independently sampled for each training step.
We set $k = \lfloor \frac{N}{6} \rfloor$ for each spectrogram.
For the transformation $T$, we use linear interpolation to make each segment have an equal unit timestep (\ie, length 1).

We adopt Tacotron 2~\cite{shen2018natural} as our backbone TTS model and denote it as Tacotron for simplicity.
Tacotron has the input format of text embedding; thus, the spectrogram inputs are not directly applicable.
To feed the warped spectrograms to the model's encoder as input, we replace the text embedding look-up table of Tacotron with a simple 1D convolutional layer.
It maps the mel-dimension to the embedding dimension of the Tacotron encoder during unsupervised training.

Other segmentation methods can be used as $S$ to generate segments instead of our proposed random segmentation, \eg, more semantically aligned segments like exact phonemes.
For example, one can adopt the Montreal Forced Alignment (MFA)~\cite{mcauliffe2017montreal} tool to extract phoneme segments using text annotations. 
However, it is not applicable in our unsupervised pre-training setting, where no text annotation is available.
Instead, one can use unsupervised pseudo phoneme segmentation~\cite{kreuk2020self}.
We empirically show that these semantic segmentation methods can improve the performance of \emph{Speech De-warping}, but the simple random segmentation is powerful enough to outperform other baselines.

\subsection{Fine-tuning: Transferring Knowledge to TTS}
After pre-training Tacotron with the pretext task, we fine-tune the model with the downstream TTS task of a target speaker.
We use a few transcribed speeches, \ie, text-audio pairs, of the target speaker to fine-tune the model.
Before starting fine-tuning, to feed texts to the model as in the original Tacotron, the 1D convolutional layer preceding the encoder is discarded, and a learnable text embedding look-up table for the target speaker's language is randomly initialized. 
With this reconfiguration, we fine-tune the TTS model for the small target speaker data.

\vspace{2mm}
\paragraph{Data Augmentation}
To further improve data efficiency during fine-tuning, we propose a simple data augmentation method called \emph{SegAug}.
During fine-tuning, we augment the training data by applying the segment-based speech warping (Equation~\ref{eqn:warping}) to the target spectrograms.
Specifically, for $S$, we use random segmentation as in the pre-training stage.
For the transformation $T$, we use linear interpolation to resize each segment of the input spectrogram along the time axis by a factor uniformly sampled from $[\frac{1}{3}, \frac{5}{3}]$.
The resulting warped spectrograms are used as the target spectrograms for training loss.
After training the model with this augmentation, we additionally train the model for a few steps without the augmentation to adapt the model to the ground truth prosody of the target speaker, \ie, a cool-down step.
Note that this augmentation in the fine-tuning stage is optional.
While our pre-training alone empirically demonstrates favorable performance, we can further improve the performance with this augmentation during fine-tuning.

% !TEX root = ../main.tex

\section{Experiments}
\label{sec:experiments}
\subsection{Experiment Setup}
\label{sec:exp_setup}

\paragraph{Dataset and Evaluation}
We use the \textit{train-clean-100} subset of the LibriTTS~\cite{zen2019libritts} dataset as the untranscribed pre-training set, which consists of 47.6 hours of speech from 247 English speakers.
We set Korean as a low-resource language and use the Korean Single speaker Speech (KSS)~\cite{park2018kss} dataset 
as our transcribed fine-tuning set.
Following \cite{chung2019semi,zhang2020unsupervised}, we define 24 minutes of speech as 1~shard of data.
Then, we construct fine-tuning datasets by randomly sampling 0.5, 1, 2, 3, 5, 8~shards from the KSS dataset.
For evaluation, we conduct both objective and subjective tests.
For the objective evaluation, we use Mel-cepstral Distortion with Dynamic time-warping (MCD-DTW) \cite{kubichek1993mel}, simply denoted as MCD. 
The objective results are reported as an average over the test set containing 571 utterances (about 22.7 minutes in total).
For the subjective evaluation, we conduct AB preference tests on 20 utterances randomly sampled from the test set.
We ask 15 native Korean raters to choose the more preferred one among two synthesized audios given the text, in terms of pronunciation, recognizability, and naturalness.

\paragraph{Implementation details}
For pre-training, we use the Adam \cite{kingma2015adam} optimizer with a learning rate of $10^{-3}$.
The models are pre-trained for 100K steps with batch size 16.
For fine-tuning, we gradually decrease the learning rate from $10^{-3}$ to $10^{-4}$ for 50K training steps with batch size 32.
Audio waveforms are down-sampled to 16~kHz, and Griffin-Lim~\cite{griffin1984signal} algorithm is used as a vocoder for fast experiment cycles.

\paragraph{Compared methods}
We use Tacotron~2~\cite{shen2018natural} as the TTS model in our experiments.
Following the naming conventions of Zhang~\etal\cite{zhang2020unsupervised} with T(acotron), we denote the model only trained with the fine-tuning data without pre-training by Tac. 
We denote two recent unsupervised pre-training methods, decoder pre-training~\cite{chung2019semi} and VQ-VAE-based pre-training~\cite{zhang2020unsupervised}, by T-Dec and T-VQ, respectively.
The model pre-trained with our \emph{Speech De-warping} is denoted by T-SD, \ie, ours.
As an upper bound of performance for the unsupervised pre-training methods, we employ the model pre-trained in a supervised manner with text annotations and denote it by T-Pho, as suggested by Zhang~\etal\cite{zhang2020unsupervised}.
In addition to the pre-training methods, we compare our data augmentation with other data augmentation methods in the fine-tuning stage.
We denote 
additive Gaussian noise-based augmentation~\cite{moell2022speech} by Gaussian,
mixup-based augmentation~\cite{guo22e_interspeech} by Mixup,
SpecAugment~\cite{park2019specaugment} by SpecAug.

\subsection{Results on Small Amount of Fine-tuning Data}
\begin{table}[]
\caption{
MCD results of several pre-training and data augmentation methods when being fine-tuned on 0.5 or 1 shard (12 or 24 minutes) of paired speech of the target speaker.
Note that T-Pho leverages text annotations in pre-training.\vspace{2mm}
}
\label{tab:main_mcd_0.5_shards}
\resizebox{\linewidth}{!}{%
\begin{tabular}{clccc}
\toprule
Augmentation
& \multicolumn{1}{l}{\multirow{2}{*}{Model}} & Supervised & \multicolumn{2}{c}{Paired data (in shards)}  \\ 
\cline{4-5}
in fine-tuning & \multicolumn{1}{l}{}                       &       pre-training              & 0.5            & 1                           \\ 
\hline
\multirow{5}{*}{No aug.}       & Tac                                        & $\times$                           & 11.98          & 12.41                       \\
                                        & T-Dec                                      & $\times$                           & 12.07          & 12.18                       \\
                                        & T-VQ                                       & $\times$                           & 11.11          & 10.41                       \\
                                        & T-SD (Ours)                  & $\times$                           & \textbf{10.79} & \textbf{10.40}              \\
                                        & T-Pho                                      & $\bigcirc$                           & 10.40          & 10.28                       \\ 
\hline
\multirow{9}{*}{With aug.}        & \multicolumn{2}{l}{Tac + Gaussian}                                      & 12.59          & 12.33                       \\
                                        & \multicolumn{2}{l}{Tac + Mixup}                                         & 12.06 & 12.04                       \\
                                        & \multicolumn{2}{l}{Tac + SpecAug}                                       & 12.29          & 10.60              \\
                                        & \multicolumn{2}{l}{Tac + SegAug}                                      & 12.19          & 10.68                       \\ 
\cdashline{2-5}[1pt/1pt]
                                        & \multicolumn{2}{l}{T-VQ + Gaussian}                                     & 10.63          & 10.40                       \\
                                        & \multicolumn{2}{l}{T-VQ + Mixup}                                        & 11.12          & 10.48                       \\
                                        & \multicolumn{2}{l}{T-VQ + SpecAug}                                       & 10.46          & 10.33                       \\
                                        & \multicolumn{2}{l}{T-VQ + SegAug}                                     & 10.41 & 10.27              \\ 
\cdashline{2-5}[1pt/1pt]
                                        & \multicolumn{2}{l}{T-SD + SegAug (Ours)}                                     & \textbf{10.28} & \textbf{10.24}              \\
\bottomrule
% \hline
\end{tabular}
}
\end{table}
%%%%%%%%%%%%%%%%%%%%%%%%%%%%%%%%%%%%%%%%%%%%%%%%%%%%%%%%%%%%%%%%%%%%%%%%%%%%%%%%%%%%%%%%

\paragraph {Objective Evaluation}
Table~\ref{tab:main_mcd_0.5_shards} presents the superior performance of the proposed methods compared to competing methods on small amounts of fine-tuning data.
Without data augmentation during fine-tuning, T-SD outperforms all unsupervised pre-training methods and Tac.
Both T-Dec~\cite{chung2019semi} and Tac, which do not have the opportunity to pre-learn a sufficient capability of attention alignment in pre-training, show similarly lower performance than the others.
In contrast, the proposed de-warping task encourages the model to learn both preliminary knowledge of attention alignment and autoregressive prediction.
When data augmentation is applied during fine-tuning, T-SD with \emph{SegAug} outperforms other combinations of pre-training and augmentation methods.
\emph{SegAug} even effectively improves the performance of other pre-training baselines and shows competitive performance compared to other augmentation methods.

\paragraph{Subjective Evaluation}
Table~\ref{tab:main_ab_test} shows the preference test results with competitive methods using 0.5 shards of fine-tuning data.
Consistent with the objective results, our methods outperform the prior art of unsupervised methods (T-VQ).
Interestingly, with our proposed data augmentation applied during fine-tuning, our whole transfer learning scheme even outperforms the supervised pre-training baseline (T-Pho).

\begin{table}[t]
\caption{
AB test results of our method over competitive baselines.
All methods use 0.5 shards (12 minutes) of fine-tuning data.\vspace{2mm}
}\label{tab:main_ab_test}
\centering
\resizebox{\linewidth}{!}{%
\begin{tabular}{lccc}
\toprule
\multirow{2}{*}{Model pair} & \multicolumn{3}{c}{Preference (\%)} \\
\cline{2-4}
    & Former & Latter & Neutral \\
\midrule
T-VQ vs. T-SD    & 15.7 & \textbf{54.7} & 29.6 \\
T-VQ vs. T-SD + SegAug    & 3.7 & \textbf{76.0} & 20.3 \\
T-Pho vs. T-SD + SegAug   & 23.3 & \textbf{44.0} & 32.7 \\
\bottomrule
\end{tabular}
}
\end{table}

\begin{figure}[t]
% \vspace{-2mm}
  \centering
  \includegraphics[width=\linewidth]{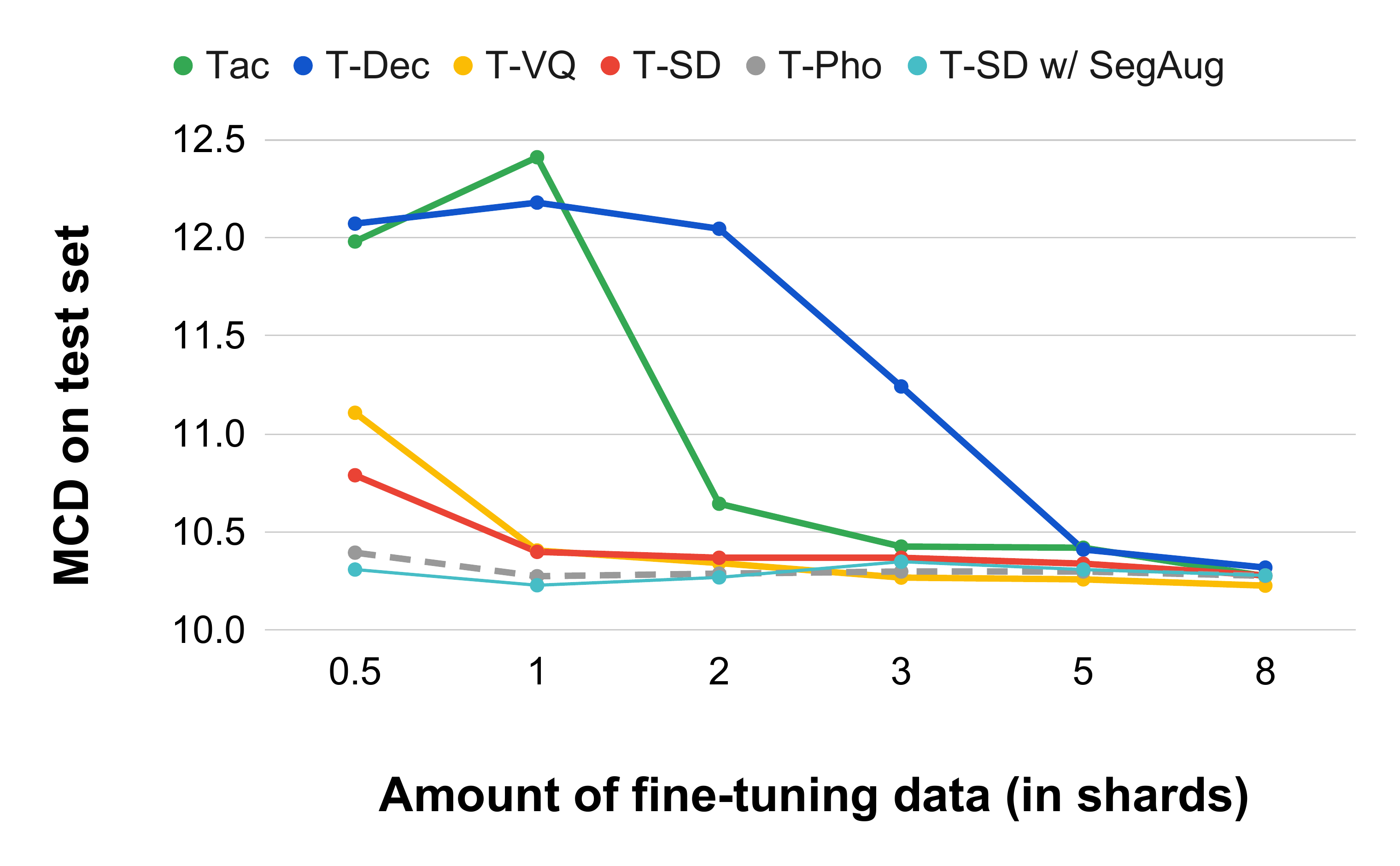}
%   \vspace{-6mm}
  \caption{
MCD results according to varying 
% for several 
amounts of paired data.
The dashed line denotes the MCD of T-Pho, which is a supervised method; thus, it can be considered a near-upper-bound performance of unsupervised pre-training methods.
  }
  \label{fig:main_mcd}
\end{figure}

\begin{table}[t]
\caption{
MCD results for the ablation study comparing our de-warping to the up-sampling pre-training.
\emph{Naive} indicates the model pre-trained with the up-sampling task.
}
\label{tab:ablation_mcd_naive}
\vspace{2mm}
% \vskip 0.15in
% \begin{center}
\centering
\resizebox{0.65\linewidth}{!}{%
\begin{tabular}{ccc} 
\toprule
% \hline
\multirow{2}{*}{Method} & \multicolumn{2}{c}{Paired data (in shards)}  \\ 
\cline{2-3}
                        & 0.5            & 1                           \\ 
\hline
Naive~                  & 11.37          & 10.88                       \\
T-SD                    & \textbf{10.79} & \textbf{10.40}              \\
\bottomrule
\end{tabular}}
\end{table}

\begin{table}
% \vspace{3mm}
\caption{
MCD results of T-VQ and our \emph{Speech De-warping} according to different segmentation methods on two fine-tuning languages.
Different and Same denote that the fine-tuning language is different or the same as the pre-training language. 
Note that phoneme segmentation requires text supervision.
}\vspace{2mm}
\label{tab:ablation_english}
% \vskip 0.15in
% \begin{center}
\centering
\resizebox{\linewidth}{!}{
\begin{tabular}{ccccc} 
\toprule
% \hline
\multirow{3}{*}{Method}      & \multicolumn{4}{c}{Paired data (in shards)}                        \\ 
\cline{2-5}
                             & \multicolumn{2}{c}{Different}   & \multicolumn{2}{c}{Same}         \\ 
\cline{2-5}
                             & 0.5            & 1              & 0.5            & 1               \\ 
\hline
T-VQ~                        & 11.11          & 10.41          & 11.85          & 10.51           \\
T-SD (Random segment)      & 10.79          & 10.40          & 11.57          & 10.63           \\
Pseudo phoneme segment& {10.56} & 
{10.38} & 11.71          & 10.69           \\
Phoneme segment             & 11.35          & 10.48          & {11.19} & {10.46}  \\
\bottomrule
% \hline
\end{tabular}
}
\end{table}

\subsection{Effects of 
% Results on Other
Different Amounts of Fine-tuning Data}

Figure~\ref{fig:main_mcd} presents the MCD evaluation results of the competing
% various
methods according to
% on
different 
% several
amounts of fine-tuning data.
Our method shows the best performance overall and is particularly better on small amounts of data.
As the amount of fine-tuning data increases, the models tend to show better performance, and the performance gaps between the methods gradually decrease.

\begin{figure*}[t]
  \centering
  \setlength\tabcolsep{0.02\linewidth}
  \begin{tabular}{ccc}
  \includegraphics[width=0.3\linewidth]{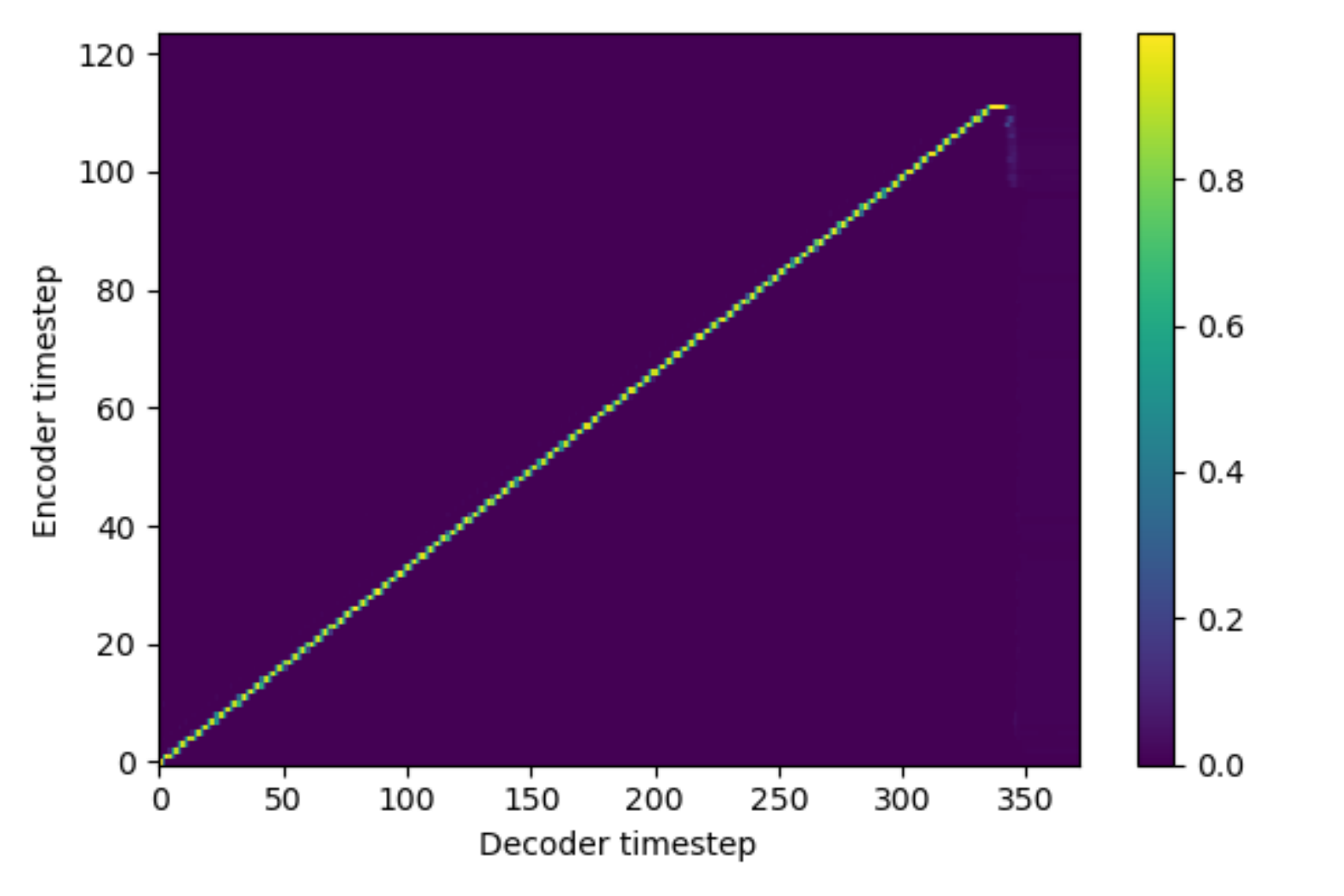} &
  \includegraphics[width=0.3\linewidth]{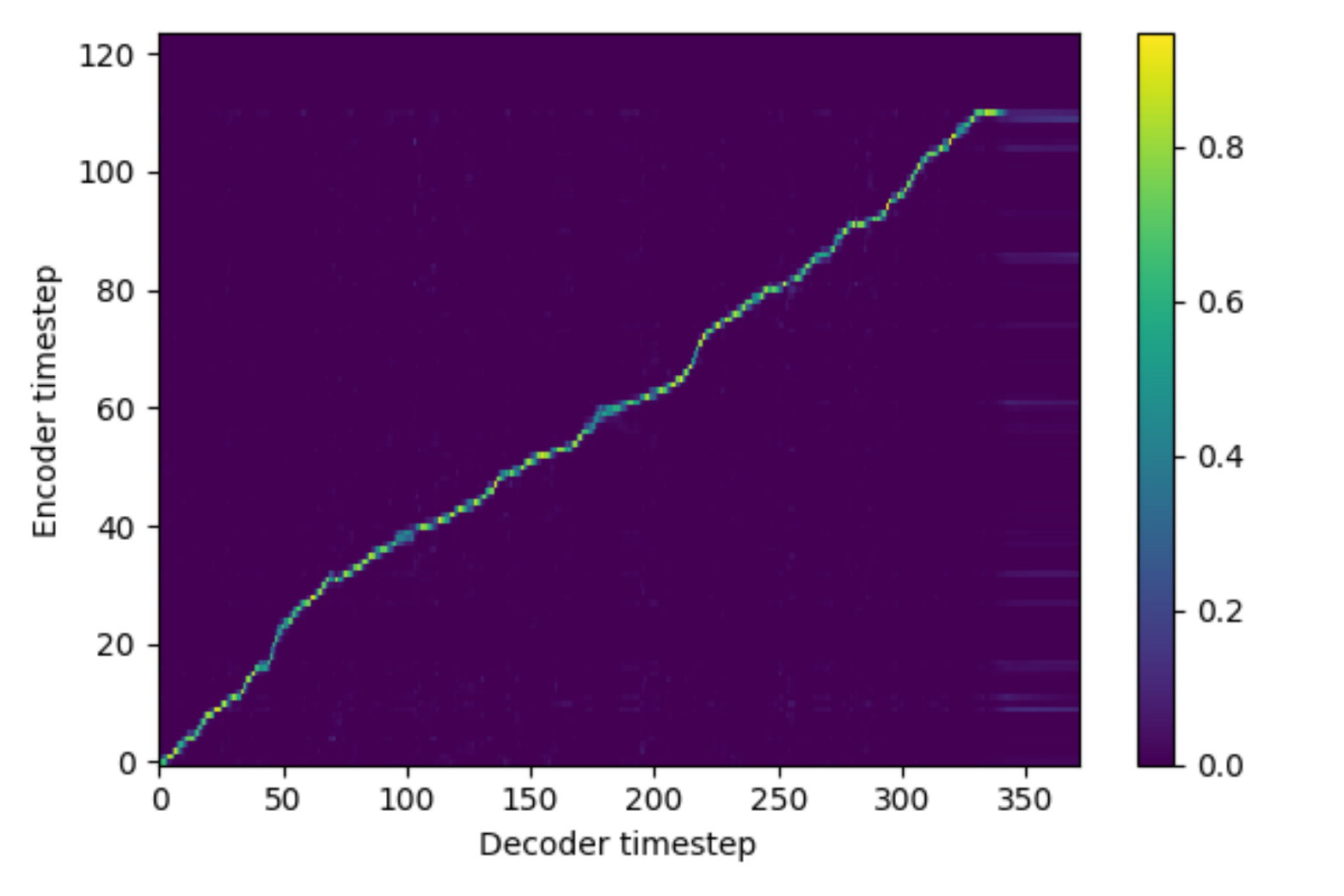} &
  \includegraphics[width=0.3\linewidth]{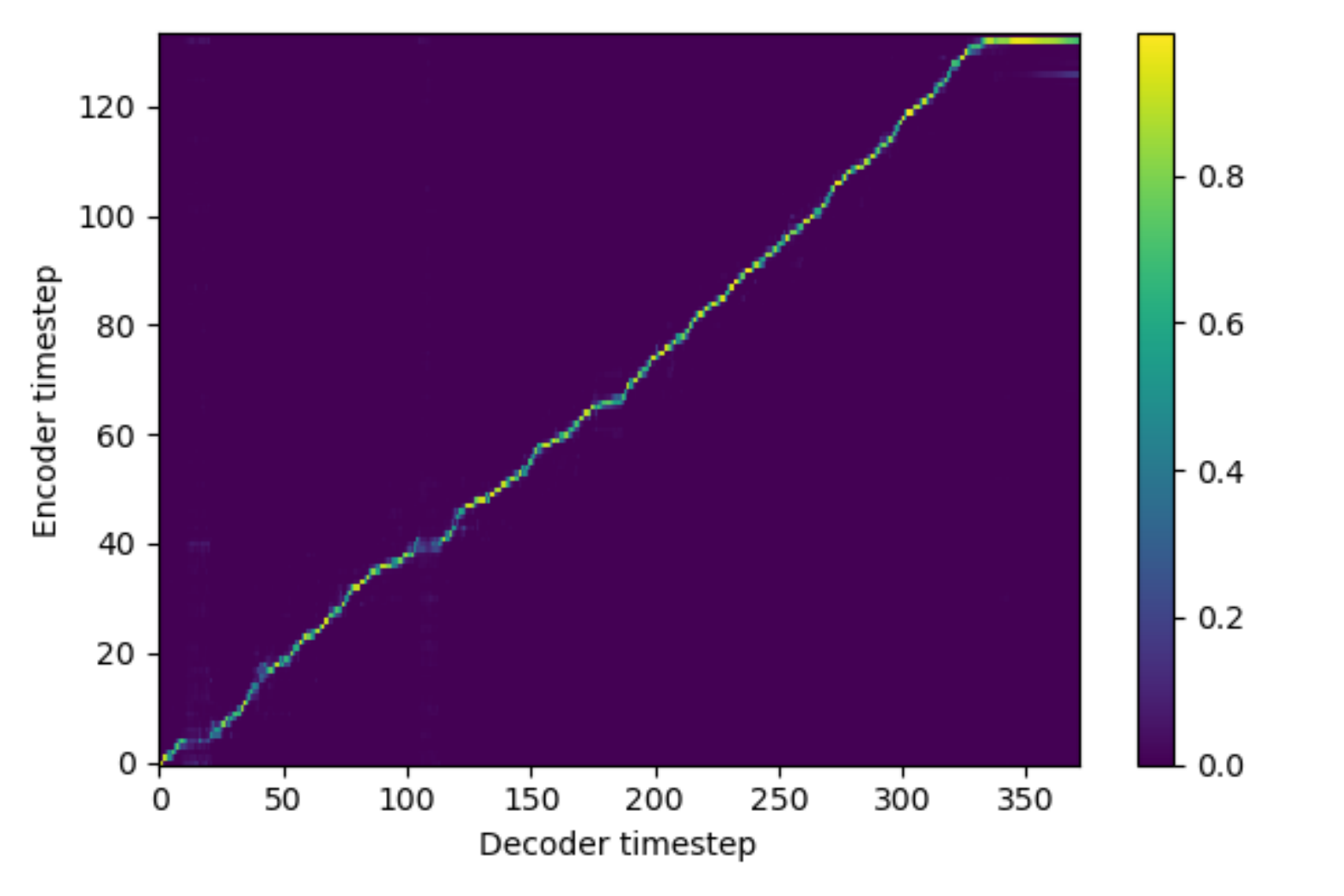}  
 \\
  (a) \emph{Naive} & (b) T-SD & (c) T-Pho\\

  \end{tabular}
  \caption{
  Examples of the learned attention alignments between input and output timesteps of the decoders of the models during pre-training.
  While \emph{Naive} induces the model to learn a linear alignment, our T-SD encourages the model to learn a non-linear alignment whose form is similar to the alignment between text and speech as in T-Pho.
  }
  \label{fig:obvious/non-obvious alignment}
\end{figure*}

\subsection{Additional Results}
\paragraph{Comparison to an upsampling pre-text task}
In our \emph{Speech De-warping}, we resize segments of different lengths into the same timestep 1 to warp the input spectrograms.
As a result, the alignment between the warped spectrogram and the original one becomes non-linear, which is analogous to 
% like 
the alignment characteristics between text and speech.
We argue that \emph{learning this monotonic yet non-linear alignment in pre-training is one of the critical factors} of our method.
To validate this argument, we introduce a control experiment with simple upsampling pre-training as a pre-text task, called \emph{Naive}, and compare it with our \emph{Speech De-warping} in Table~\ref{tab:ablation_mcd_naive}.
Specifically, in \emph{Naive}, 
instead of using the segment-wise warping to warp the spectrogram, we downsample the whole spectrogram by a single scale factor of $\tfrac{1}{6}$
using linear interpolation along the time axis.
Thereby, the model with \emph{Naive} learns a linear alignment between the uniformly downsampled spectrograms and the original spectrograms.
Figure~\ref{fig:obvious/non-obvious alignment} presents examples of attention alignments learned during pre-training.
The superior performance of T-SD compared to \emph{Naive} in Table~\ref{tab:ablation_mcd_naive} verifies that learning a \emph{monotonic and non-linear alignment} benefits our \emph{Speech De-warping}.
Note that \emph{Naive} performs better than Tac and T-Dec in Table~\ref{tab:main_mcd_0.5_shards}, which demonstrates the effectiveness of learning \emph{monotonic alignment} through the upsampling task itself.

\paragraph{Effect of Heterogeneous Languages in Fine-tuning}
We investigate the effect of using different or the same languages between pre-training and fine-tuning steps, which is the main scope of this work.
As described, we use English as pre-training data.
For the same language scenario, called \emph{Same}, we use the LJspeech~\cite{ljspeech17} dataset (English) for fine-tuning data.
The different language scenario follows the same setup described in Sec.~\ref{sec:exp_setup}, called \emph{Different}.
We compare our T-SD with T-VQ to show the algorithmic behavioral differences.
Table~\ref{tab:ablation_english} shows the performance of T-SD is overall similar to T-VQ when the language between pre-training and fine-tuning is unchanged, \ie, \emph{Same}.
However, as shown in the \emph{Different} columns, T-SD is more robust against overfitting to the pre-training language than T-VQ.
We conjecture this is because the burden to memorize acoustic features of the pre-training language is less for our method since some language-specific acoustic information is already given as input for de-warping.

\paragraph{Effect of Segmentation Methods}
We investigate the effect of different segmentation methods for the segment-based speech warping in \emph{Speech De-warping}.
In addition to the random segmentation used in our T-SD, we compare with the phoneme segmentation by using the MFA tool~\cite{mcauliffe2017montreal}, which requires text supervision, and the pseudo phoneme segmentation by using the unsupervised phoneme segmentation model~\cite{kreuk2020self}.
As shown in Table~\ref{tab:ablation_english}, the performance of \emph{Speech De-warping} can be boosted by using semantically meaningful segmentation obtained from external models.
The phoneme segmentation shows the best performance when the fine-tuning language is the same as the pre-training language and the worst when the fine-tuning language is unseen during pre-training.
The phoneme segmentation of a specific language induces the alignment of the warped spectrograms and original spectrograms to be very similar to the alignment between text and speech of that language in pre-training.
This behavior can lead to overfitting to the specific language used in pre-training.

% !TEX root = ../main.tex

\section{Conclusion}
\label{sec:conclusion}
We propose an unsupervised pre-training method and a data augmentation method for training TTS models with limited amounts of text-annotated speech data.
Our pre-training method enables us to build a TTS system for a low-resource language by leveraging a large-scale and untranscribed speech dataset that can be easily collected.
The proposed data augmentation technique can be used to further improve such data efficiency.
Our comprehensive experiments show the superior
performance of the proposed methods compared to various competing pre-training and data augmentation methods.
We empirically demonstrate that learning a non-linear alignment during pre-training of the model is beneficial in TTS compared to learning a linear alignment.
We show that our pre-training method can achieve better performance by using external models for segmentation.

\paragraph{Acknowledgments}
T.-H. Oh was partially supported by Institute of Information \& communications Technology Planning \& Evaluation (IITP) grant funded by the Korea government (MSIT) (No.2021-0-02068, Artificial Intelligence Innovation Hub; No.2022-0-00124, Development of Artificial Intelligence Technology for Self-Improving Competency-Aware Learning Capabilities; No. 2019-0-01906, Artificial Intelligence Graduate School Program(POSTECH)).

% \begingroup
%     \fontsize{10pt}{10pt}\selectfont
%         methods
% \endgroup

% \vfill\pagebreak
% \section{REFERENCES}
% \label{sec:refs}

% References should be produced using the bibtex program from suitable
% BiBTeX files (here: strings, refs, manuals). The IEEEbib.bst bibliography
% style file from IEEE produces unsorted bibliography list.
% -------------------------------------------------------------------------
\bibliographystyle{IEEEbib}
\bibliography{main.bbl}

\end{document}